\documentclass[prb,twocolumn,aps,superscriptaddress,showpacs,floatfix]{revtex4-1}
\usepackage{amsbsy,amssymb,amsmath,bm}
\usepackage{graphicx,color,epsfig,rotate}
\usepackage{xspace,units}
\usepackage{textcomp}		
\usepackage{epstopdf}
\usepackage{tabularx}

\usepackage{multirow}		
\usepackage{dcolumn}							
\newcolumntype{d}[1]{D{.}{.}{#1}}
 
\renewcommand\thesection{\Roman{section}}
\renewcommand\thesubsection{\Alph{subsection}}

\newcommand{\Gd}{Gd(NO$_3$)$_3$\:}
\newcommand{\Y}{Y(NO$_3$)$_3$\:}
\newcommand{\Nd}{\mbox{Nd-Fe-B}\:}

\begin{document}
\title{Neutron imaging of liquid-liquid systems containing paramagnetic salt solutions}

\author{T. A. Butcher}
\email{tbutcher@tcd.ie}
\affiliation{School of Physics and CRANN, Trinity College, Dublin 2, Ireland}

\author{G. J. M. Formon}
\affiliation{Universit\'{e} de Strasbourg, CNRS, ISIS UMR 7006, 67000 Strasbourg, France}

\author{P. Dunne}
\affiliation{Universit\'{e} de Strasbourg, CNRS, ISIS UMR 7006, 67000 Strasbourg, France}

\author{T. M. Hermans}
\affiliation{Universit\'{e} de Strasbourg, CNRS, ISIS UMR 7006, 67000 Strasbourg, France}

\author{F. Ott}
\affiliation{Laboratoire L\'eon Brillouin (CEA-CNRS), Universit\'{e} Paris-Saclay, CEA-Saclay, 91191 Gif-sur-Yvette, France}

\author{L. Noirez}
\affiliation{Laboratoire L\'eon Brillouin (CEA-CNRS), Universit\'{e} Paris-Saclay, CEA-Saclay, 91191 Gif-sur-Yvette, France}

\author{J. M. D. Coey}
\affiliation{School of Physics and CRANN, Trinity College, Dublin 2, Ireland}

\date{December 19, 2019}

\begin{abstract}

\noindent The method of neutron imaging was adopted to map the concentration evolution of aqueous paramagnetic \Gd solutions. Magnetic manipulation of the paramagnetic liquid within a miscible nonmagnetic liquid is possible by countering density-difference driven convection.
The formation of salt fingers caused by double-diffusive convection in a liquid-liquid system of \Gd and \Y solutions can be prevented by the magnetic field gradient force. 

\end{abstract}


\maketitle

Paramagnetic liquids are created by dissolving salts containing transition-metal or rare-earth ions in a solvent \cite{andres_1976}. Magnetic levitation of objects immersed in paramagnetic liquids has been used for magnetohydrostatic separation since the 1960s\cite{andres_1966} and nowadays finds application in biotechnology\cite{turker_2018}. Exposing a paramagnetic solution to an inhomogeneous magnetic field gives rise to the magnetic field gradient force\cite{mutschke_2010, dunne_2011, dunne_2012, remark_kelvin}:

\begin{equation}
\mathbf{F}_{\nabla B} = \frac{\chi}{2\mu_0 } \nabla \mathbf{B}^2.
\end{equation}

This expression relates the force density to the magnetic susceptibility of the solution $\chi$, the magnetic flux density $B$, and the permeability of free space $\mu_0$. 

It is possible to trap aqueous paramagnetic salt solutions in the magnetic field gradient of a magnetized iron wire\cite{coey_2009}. Convection from these paramagnetic liquid tubes is inhibited by the magnetic field gradient force, although mixing by diffusion still prevails on a long time scale. A magnetic field gradient can also initiate magnetothermal convection in a paramagnetic fluid\cite{braithwaite_1991, rodrigues_2019}.

At present, the possibility of extracting paramagnetic ions from a homogeneous aqueous solution with an inhomogeneous magnetic field is garnering significant research interest \cite{eckert_2012, eckert_2014, rodrigues_2017, eckert_2017, franczak_2016, kolczyk_2016, kolczyk_2019}. In response to the recent rare-earth crisis, this activity has been spurred by the idea of magnetically separating rare-earth ions, which was originally explored by Noddack et al. in the 1950s \cite{noddack_1952, noddack_1955, noddack_1958}. Recent studies used Mach-Zehnder interferometers to relate changes in the refractive index to an enrichment of magnetic ions underneath a permanent magnet\cite{eckert_2012,eckert_2014,rodrigues_2017,eckert_2017}. The two most recent of these showed that the observed magnetic enrichment is evaporation-assisted\cite{rodrigues_2017, eckert_2017}. According to Lei et al. \cite{eckert_2017}, the heightened ion concentration \mbox{($\leq2$\%} bulk concentration) in the evaporation layer is maintained by the magnetic field gradient. This results in a modest long-lived paramagnetic ion enrichment underneath the magnet. The magnetic field gradient force pales in comparison with the force governing diffusion and is unable to appreciably influence the motion of individual ions on these grounds ($R T \nabla
c \approx 10^{10}\,\mathrm{N}/\mathrm{m}^3 \gg F_{\nabla B} \approx 10^{4}\, \mathrm{N}/\mathrm{m}^3 $; \mbox{$R$: gas constant}, $T$: room temperature, and $\nabla c$: concentration gradient)\cite{rodrigues_2019}.

\begin{figure}
	\centering
	\includegraphics[width=0.99\columnwidth]{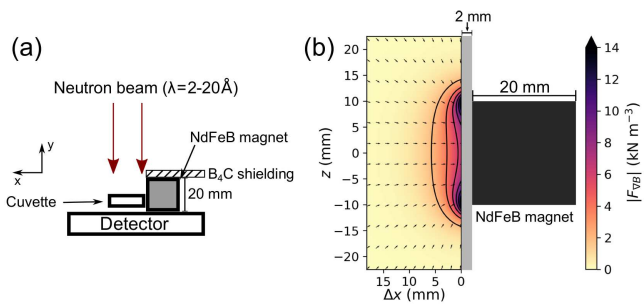}
	\caption{(a) Sketch of the experimental setup (top view). (b)~Calculated magnetic field gradient force distribution in the cuvette for a 1\,M aqueous \Gd solution in the field of a uniformly magnetized 20\,mm \Nd cube.}
	\label{fig:sketch}
\end{figure}

In this study, we use neutron imaging to track the concentration distribution of aqueous paramagnetic gadolinium(III) nitrate (\Gd) solutions in a liquid-liquid system with a miscible nonmagnetic counterpart. This direct method consists of measuring the attenuation of a white neutron beam on passing through a sample and has potential for applications in a variety of fields\cite{kardjilov_2011, perfect_2014, burca_2018}. Neutrons interact with the nuclei of the sample, which makes the measurement element-specific, allowing the direct study of liquids that are both miscible and visually indistinguishable under normal conditions. Gd is the element with the highest neutron absorption cross section (\mbox{$\sigma_a=46$ $700$\,barn}  for thermal neutrons\cite{mastromarco_2019}). In addition, Gd$^{3+}$ possesses a large magnetic moment of 7\,$\mu_B$ by virtue of unpaired 4f electrons. Consequently, solutions of \Gd are ideal candidates for neutron imaging of paramagnetic solutions, enabling the direct observation of their response to magnetic fields. Recent advances in detector systems have provided the means for neutron imaging with both high spatial and temporal resolutions \cite{brenizer_2013, trtik_2016, zboray_2019}. Here, we monitor the interplay of convection, magnetic field gradient force, and diffusion by variations in the neutron transmission profile.\\
\indent Neutron imaging experiments were carried out at the IMAGINE station\cite{ott_2015} located in the neutron guide hall of the Orph\'{e}e reactor at the Laboratoire L\'{e}on Brillouin just before its final shutdown. The spectrum of the white neutron beam contained cold neutrons (\mbox{$\lambda$ = 2-20\,\AA}), which emerged from a 10\,mm pinhole and travelled 2.5\,m to the detector where the neutron flux was $2\times 10^7\,\mathrm{cm}^{-2}\, \mathrm{s}^{-1}$. The detection system consisted of a 50\,\textmu m thick $^6$LiF/ZnS scintillator, with a resolution of 18\,\textmu m/pixel, coupled to an sCMOS camera. Recorded images of 2560$\times$2160 pixels correspond to a field of view of $46\times39$\,mm$^2$. All images shown here were obtained with an acquisition time of 60\,s. The spatial resolution is on the order of 50\,\textmu m.\\
\indent Quartz cuvettes with path lengths of 1\,mm were filled with the liquid solutions and placed 5\,mm in front of the detector (see sketch in Fig.~\ref{fig:sketch}(a)). The outside dimensions of the cuvettes were \mbox{40\,mm $\times$ 23.6\,mm $\times$ 3.5\,mm} \mbox{(height $\times$ width $\times$ depth)}. Incoherent scattering by water molecules (\mbox{$\sigma_{inc} = 160.5$\,barn}) was minimized by dissolving salts in D$_2$O (\mbox{$\sigma_{inc} = 4.1$\,barn}). For maximum contrast, the analyzed paramagnetic salt solutions were restricted to colorless and transparent \Gd solutions. The neutron absorption cross section of Gd dwarfs the scattering cross section of D$_2$O (\mbox{$\sigma_s=19.5$\,barn}). Thus, effects of parasitic scattering on the final signal are expected to be weak and it is unnecessary to employ a scattering correction algorithm\cite{kardjilov_2005}. For the study of magnetic effects, a cube-shaped \Nd permanent magnet of side length 20\,mm was placed adjacent to the cuvettes (2\,mm from the solution within) and shielded from the neutron beam with a boron carbide sheet. The horizontal magnetic field was $B=0.45$\,T at the surface of the magnet and $B=0.13\,$T at a distance of 5\,mm.

An empty beam was recorded during each measurement session. This was necessary for normalization to the intensity of the white beam $I_0$. Furthermore, the electronic noise $I_{df}$ was subtracted from the image to obtain the transmittance:

\begin{equation}
\label{eq:trans}
T=\frac{I - I_{df}}{I_0-I_{df}}.  
\end{equation}

The final step of the image processing was the removal of noisy pixels by using an outlier filter.

\begin{figure}
	\centering
	\includegraphics[width=0.85\columnwidth]{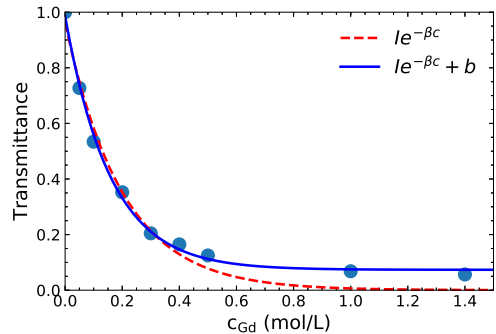}
	\caption{Gd$^{3+}$ concentration calibration curve in a 1\,mm quartz cuvette. Transmittance values were normalized to that of a D$_2$O filled cuvette and follow the Beer-Lambert law (broken line) up to 0.4\,M. An offset \mbox{$b = 0.07$} (solid line) is needed at higher concentration.}
	\label{fig:cal}
\end{figure}

The Beer-Lambert law describes the attenuation of the neutron beam by the Gd$^{3+}$ ions in D$_2$O:

\begin{equation}
I = I_0 \, e^{-\epsilon c l},
\end{equation}

\noindent with the molar neutron absorption coefficient~$\epsilon$, the Gd$^{3+}$ concentration~$c$ and the sample thickness~$l$. Strictly speaking, $\epsilon$ depends on the neutron energy and the assumption of a single value for a polychromatic neutron beam is a simplification. This approximation is not a concern, considering the fact that the neutron wavelengths constituting the beam (\mbox{2-20\,\AA}) lie within one order of magnitude of each other. 
A calibration of the transmitted intensity to the Gd$^{3+}$ concentration was performed by recording images of solutions in 1\,mm path length cuvettes (see Fig.~\ref{fig:cal}). The attenuation follows the Beer-Lambert law up to a concentration of about 0.4\,M, when the beam is almost completely absorbed and the transmitted intensity originates predominately from incoherent scattering.
An offset exponential fit with an extra variable (\mbox{$b = 0.07$}) captures the behavior, but quantitative statements cannot be readily made at concentrations higher than 0.5\,M. The latest development of black body correction\cite{boillat_2018, carminati_2019} opens the possibility to quantify the contribution of background and sample scattering to the transmittance, but it would require a black body grid. 

The magnetic susceptibility of a 1\,M \Gd heavy water solution (\mbox{$\chi_{1\mathrm{M}} = 322 \times 10^{-6}$}) is the sum of the diamagnetic D$_2$O contribution (\mbox{$\chi_{\mathrm{D}_2\mathrm{O}} = -8 \times 10^{-6}$}) and the paramagnetic \mbox{Curie-law} contribution of the Gd$^{3+}$ ions\cite{coey_2009}. This value and the magnetic field distribution of the \Nd magnet allow the computation of the magnetic field gradient force in the vicinity of the magnet (see Fig.~\ref{fig:sketch}(b)). The magnetic field was calculated by approximating the magnet as two uniform sheets of magnetic charge \cite{furlani_2001}.

\begin{figure}
	\centering
	\includegraphics[width=0.99\columnwidth]{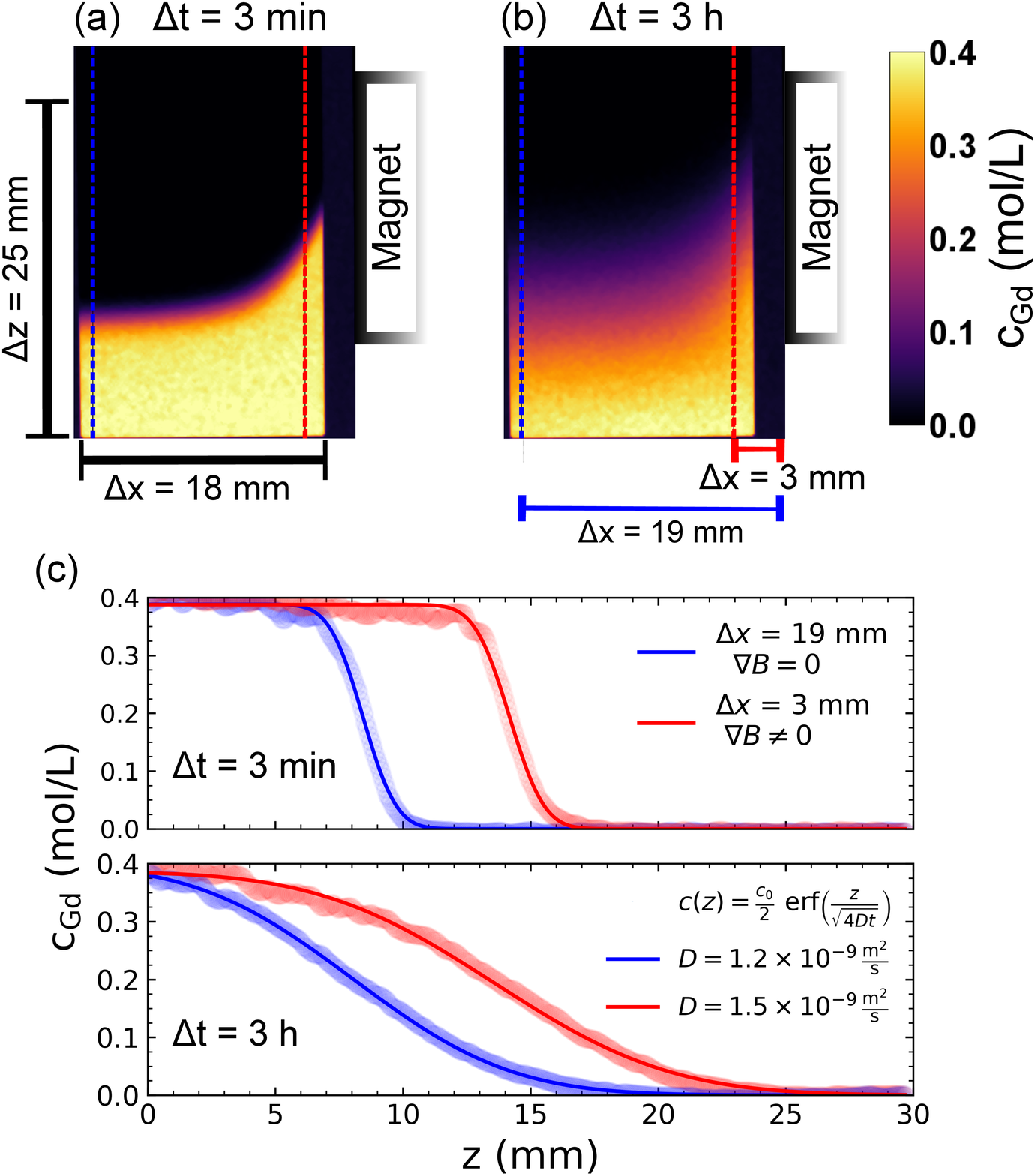}
	\caption{Neutron image of a 1\,mm path length quartz cuvette with 100\,\textmu L 0.4\,M \Gd solution overlain with 400\,\textmu L of D$_2$O. A 20\,mm magnet cube to the right skews the Gd$^{3+}$ concentration profile. (a) After 3\,min (b) after 3\,h. (c) Fits of Eq.~(4) to the vertical cross sections (broken lines in neutron images) of the concentration profiles 3\,mm ($\nabla B \neq 0$) and 19\,mm ($\nabla B = 0$) away from the magnet show good agreement and the diffusion coefficient $D$ can be obtained.}
	\label{fig:skew}
\end{figure}

In the case of an inhomogeneous solution comprising a paramagnetic and nonmagnetic component, a magnetic field gradient orthogonal to the concentration gradient alters the equilibrium state\cite{mutschke_2010}. The \Gd solution climbs up the side of the cuvette until the balance between buoyancy (\mbox{$F_g = \Delta \rho g$}) and magnetic field gradient forces is re-established. This can be seen in Fig. \ref{fig:skew}. Here, 100\,\textmu L of 0.4\,M \Gd solution (\mbox{$\rho = 1180\,\mathrm{kg}\,\mathrm{m^{-3}}$})at the bottom of a 1\,mm path length cuvette was covered with 400\,\textmu L D$_2$O (\mbox{$\rho = 1110\,\mathrm{kg}\,\mathrm{m^{-3}}$}). A magnet was placed at the side and the diffusion of the \Gd was monitored for 3\,h. The magnetic field gradient draws the \Gd solution towards the magnet, although homogenization by diffusion continues in its presence. An estimate for the diffusion coefficient $D$ of 0.4\,M \Gd  in D$_2$O can be obtained from the vertical concentration profile by a fit with the solution of the one-dimensional diffusion equation (see Fig.~\ref{fig:skew}(c)):

\begin{equation}
c(z, t) = \frac{c_0}{2} \, \mathrm{erf}\!\left(\frac{z}{\sqrt{4Dt}}\right),
\end{equation}

with Gd$^{3+}$ starting concentration $c_0$. The value of \mbox{$D=1.2\times 10^{-9}\,$m$^2$s$^{-1}$} obtained for the nonmagnetized region after 3\,h is reasonable for rare-earth ions in water\cite{cussler_2009}. However, this value should be treated with caution, as the initial interface was smeared by introducing the liquids into the cuvette before the onset of diffusion. The diffusion coefficient from the fit for the magnetized region is higher at \mbox{$D=1.5\times 10^{-9}$\,m$^2$s$^{-1}$}, but the one-dimensional expression does not account for horizontal diffusion from the warped concentration profile. 

\begin{figure}
	\centering
	\includegraphics[width=0.999\columnwidth]{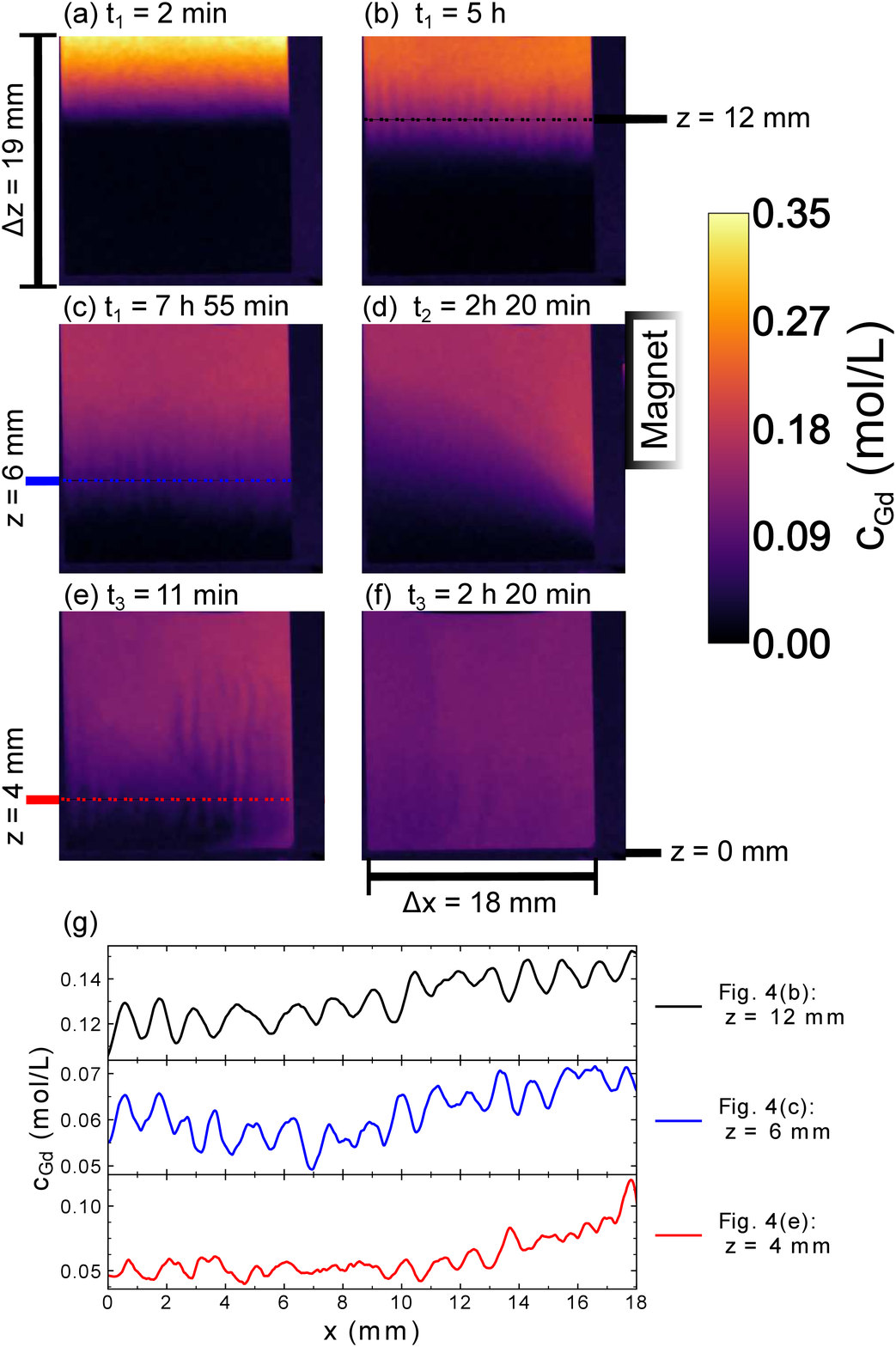}
	\caption{(a)-(f) Neutron images of 100\,\textmu L 0.4\,M \Gd above 300\,\textmu L 1.3\,M \Y solution ($\Delta \rho = 140\,\mathrm{kg}\,\mathrm{m^{-3}}$) in a 1\,mm path length quartz cuvette. The view is restricted to the area below the surface in the vicinity of the liquid-liquid interface. (a) 2 min after the \Gd solution is suspended above the \Y surface. (b)-(c) Double-diffusion imposes Gd salt fingers which protrude into the \Y solution after 90\,min and begin to sink due to the loss of buoyancy. The fingers have a width of 1.2\,mm and persist for over 8\,h (Multimedia view). (d) A cubic 20\,mm magnet at the side of the cuvette halts the instability growth and destroys the stratification instantly ($t_2$: time since magnetization; Multimedia view). (e)-(f)  Once the magnet is removed, the control over the \Gd is relinquished and it fans out. The system snaps back into the stratified state in less than 10\,min and the cascading salt fingers homogenize the mixture after 2\,h ($t_3$: time since removal of magnet; Multimedia view). (g) Horizontal cross sections (broken lines in the neutron images) of the salt fingers in (b), (c), and (e) show a periodic variation of the Gd$^{3+}$ concentration by $\approx$0.02\,mol\,L$^{-1}$. The plotted data was smoothed with a Savitzky-Golay filter.}
	\label{fig:finger}
\end{figure}

The density difference between the \Gd solution and the nonmagnetic liquid can be adjusted by addition of Yttrium(III) nitrate (\Y), which is transparent to neutrons (\mbox{$\sigma_a=7.0$\,barn}\cite{varley_1992}), to the D$_2$O. 
Decreasing the density difference leads to more vigorous magnetically induced migration and facilitates magnetic confinement. If the density of the Gd solution is higher than that of the Y solution, the removal of the magnet before homogenization has taken place prompts a buoyancy-driven Rayleigh-Taylor instability\cite{huang_2007, tsiklashvili_2012} and the Gd solution plunges to the bottom of the cuvette in a matter of seconds. A different situation arises when the density difference is inverted and the Gd solution floats above the Y solution. To investigate this, 100\textmu L \,0.4\,M \Gd solution was injected on top of 300\,\textmu L 1.3\,M \Y solution ($\rho = 1320\,\mathrm{kg}\,\mathrm{m^{-3}}$) in a cuvette (see Fig.~\ref{fig:finger}(a)). After 1\,h the system was beset by a salt-fingering instability due to double diffusive convection\cite{turner_1964, stern_1969} (see Fig.~\ref{fig:finger}~(b)-(c) and animations for greater visibility). This phenomenon is encountered at the interface of solutions that diffuse into each other at unequal rates. The diffusivity of Y$^{3+}$ in the 1.3\,M solution exceeds that of the Gd in the 0.4\,M solution. It follows that \Y will diffuse laterally into small portions of \Gd solution that cross the interface. The increase in density due to the gained \Y makes the \Gd solution plummet in form of 1.2\,mm wide fingers (see cross sections in Fig.~\ref{fig:finger}(g)), which continue to leech \Y from their surroundings during their descent. These transport the \Gd advectively, two orders of magnitude faster than regular diffusion and trigger a stratification with neighboring fingers that rise thanks to the buoyancy acquired by the loss of \Y. Hence, the usually stabilizing factor of diffusion can destabilize a system in which the density decreases upwards. The stratification persists for over 8\,h (see Fig.~\ref{fig:finger}(c) and animations in supplementary material). Viscous friction between the liquid and the cuvette walls plays a role in the horizontal scale of the individual fingers, which is inversely proportional to the distance between the cell walls\cite{taylor_1986}. Thus, a horizontal expansion of the fingers beyond the gap width is achievable in thin cuvettes. A magnet next to the cuvette erases the stratification and restabilizes the system by capturing the paramagnetic solution (see Fig.~\ref{fig:finger}(d)). This does not reverse the mixing that has occurred and the Gd$^{3+}$ ions can be seen to continuously diffuse into the \Y solution. The magnetic field gradient merely prevents the collapse of the liquid-liquid interface. Nonetheless, the system undergoes an immediate change upon its withdrawal (see Fig.~\ref{fig:finger}(e)). Bereft of the confining magnetic field gradient force, the boundary between the solutions is once again disrupted. The ensuing release of the paramagnetic liquid is accompanied by convective mixing of the solutions amidst which the salt fingering instability can be witnessed anew. After two hours the system equilibrates as homogenization sets in (see Fig.~\ref{fig:finger}(f)).


In conclusion, neutron imaging is a viable method for capturing quasi two-dimensional convective and diffusive processes in solutions containing Gd$^{3+}$ ions. A pre-existing concentration of paramagnetic fluid in some region can be redistributed within a miscible liquid by the magnetic field gradient force, which counteracts density-difference driven convection. Furthermore, double-diffusive convection in the system of magnetic \Gd and nonmagnetic \Y salt solutions is suppressed. This manifests itself in the stratification by salt fingers when the magnet is absent. The implication of this is of great importance for the development of the magnetic separation of rare-earth ions, as even minor differences in diffusivity can precipitate salt fingering instabilities. If left unchecked, these will mix the separated solutions. A prerequisite for the generation of the liquid-liquid interface is a driving force that creates and preserves the concentration gradient of the paramagnetic ions. The magnetic field gradient is then able to bestow stability upon the system. Driving forces can range from the weak factor of evaporation to the more substantial injection of electrochemical energy, which can drive convection \cite{mutschke_2010, dunne_2011, dunne_2012}. In view of improvements in both imaging instrumentation and available neutron flux, higher resolution and frame rates are expected to improve the neutron imaging of hydrodynamic processes in the future\cite{morgano_2015, trtik_2016_1}. This may prove valuable for the analysis of ions in solutions.

~\\
\indent See supplementary material for time sequenced images of liquid-liquid systems containing \Gd and \Y solutions: undisturbed double-diffusive convection and magnetic confinement of a \Gd drop. 

~\\
\indent
This work forms part of the MAMI project, which is an Innovative Training Network funded by the European Union's Horizon 2020 research and innovation program under grant agreement No. 766007. 

\bibliography{imagine}

\begin{thebibliography}{41}%
\makeatletter
\providecommand \@ifxundefined [1]{%
 \@ifx{#1\undefined}
}%
\providecommand \@ifnum [1]{%
 \ifnum #1\expandafter \@firstoftwo
 \else \expandafter \@secondoftwo
 \fi
}%
\providecommand \@ifx [1]{%
 \ifx #1\expandafter \@firstoftwo
 \else \expandafter \@secondoftwo
 \fi
}%
\providecommand \natexlab [1]{#1}%
\providecommand \enquote  [1]{``#1''}%
\providecommand \bibnamefont  [1]{#1}%
\providecommand \bibfnamefont [1]{#1}%
\providecommand \citenamefont [1]{#1}%
\providecommand \href@noop [0]{\@secondoftwo}%
\providecommand \href [0]{\begingroup \@sanitize@url \@href}%
\providecommand \@href[1]{\@@startlink{#1}\@@href}%
\providecommand \@@href[1]{\endgroup#1\@@endlink}%
\providecommand \@sanitize@url [0]{\catcode `\\12\catcode `\$12\catcode
  `\&12\catcode `\#12\catcode `\^12\catcode `\_12\catcode `\%12\relax}%
\providecommand \@@startlink[1]{}%
\providecommand \@@endlink[0]{}%
\providecommand \url  [0]{\begingroup\@sanitize@url \@url }%
\providecommand \@url [1]{\endgroup\@href {#1}{\urlprefix }}%
\providecommand \urlprefix  [0]{URL }%
\providecommand \Eprint [0]{\href }%
\providecommand \doibase [0]{http://dx.doi.org/}%
\providecommand \selectlanguage [0]{\@gobble}%
\providecommand \bibinfo  [0]{\@secondoftwo}%
\providecommand \bibfield  [0]{\@secondoftwo}%
\providecommand \translation [1]{[#1]}%
\providecommand \BibitemOpen [0]{}%
\providecommand \bibitemStop [0]{}%
\providecommand \bibitemNoStop [0]{.\EOS\space}%
\providecommand \EOS [0]{\spacefactor3000\relax}%
\providecommand \BibitemShut  [1]{\csname bibitem#1\endcsname}%
\let\auto@bib@innerbib\@empty
\bibitem [{\citenamefont {Andres}(1976)}]{andres_1976}%
  \BibitemOpen
  \bibfield  {author} {\bibinfo {author} {\bibfnamefont {U.}~\bibnamefont
  {Andres}},\ }\href {\doibase https://doi.org/10.1016/0025-5416(76)90014-8}
  {\bibfield  {journal} {\bibinfo  {journal} {Mater. Sci. Eng.}\ }\textbf
  {\bibinfo {volume} {26}},\ \bibinfo {pages} {269 } (\bibinfo {year}
  {1976})}\BibitemShut {NoStop}%
\bibitem [{\citenamefont {Andres}\ \emph {et~al.}(1966)\citenamefont {Andres},
  \citenamefont {Bunin},\ and\ \citenamefont {Gil}}]{andres_1966}%
  \BibitemOpen
  \bibfield  {author} {\bibinfo {author} {\bibfnamefont {U.~T.}\ \bibnamefont
  {Andres}}, \bibinfo {author} {\bibfnamefont {G.~M.}\ \bibnamefont {Bunin}}, \
  and\ \bibinfo {author} {\bibfnamefont {B.~B.}\ \bibnamefont {Gil}},\ }\href
  {\doibase 10.1007/BF00914718} {\bibfield  {journal} {\bibinfo  {journal} {J.
  Appl. Mech. Tech. Phys.}\ }\textbf {\bibinfo {volume} {7}},\ \bibinfo {pages}
  {109} (\bibinfo {year} {1966})}\BibitemShut {NoStop}%
\bibitem [{\citenamefont {Turker}\ and\ \citenamefont
  {Arslan-Yildiz}(2018)}]{turker_2018}%
  \BibitemOpen
  \bibfield  {author} {\bibinfo {author} {\bibfnamefont {E.}~\bibnamefont
  {Turker}}\ and\ \bibinfo {author} {\bibfnamefont {A.}~\bibnamefont
  {Arslan-Yildiz}},\ }\href {\doibase 10.1021/acsbiomaterials.7b00700}
  {\bibfield  {journal} {\bibinfo  {journal} {ACS Biomater. Sci. Eng.}\
  }\textbf {\bibinfo {volume} {4}},\ \bibinfo {pages} {787} (\bibinfo {year}
  {2018})}\BibitemShut {NoStop}%
\bibitem [{\citenamefont {Mutschke}\ \emph {et~al.}(2010)\citenamefont
  {Mutschke}, \citenamefont {Tschulik}, \citenamefont {Weier}, \citenamefont
  {Uhlemann}, \citenamefont {Bund},\ and\ \citenamefont
  {Fr{\"o}hlich}}]{mutschke_2010}%
  \BibitemOpen
  \bibfield  {author} {\bibinfo {author} {\bibfnamefont {G.}~\bibnamefont
  {Mutschke}}, \bibinfo {author} {\bibfnamefont {K.}~\bibnamefont {Tschulik}},
  \bibinfo {author} {\bibfnamefont {T.}~\bibnamefont {Weier}}, \bibinfo
  {author} {\bibfnamefont {M.}~\bibnamefont {Uhlemann}}, \bibinfo {author}
  {\bibfnamefont {A.}~\bibnamefont {Bund}}, \ and\ \bibinfo {author}
  {\bibfnamefont {J.}~\bibnamefont {Fr{\"o}hlich}},\ }\href {\doibase
  https://doi.org/10.1016/j.electacta.2010.08.046} {\bibfield  {journal}
  {\bibinfo  {journal} {Electrochim. Acta}\ }\textbf {\bibinfo {volume} {55}},\
  \bibinfo {pages} {9060 } (\bibinfo {year} {2010})}\BibitemShut {NoStop}%
\bibitem [{\citenamefont {Dunne}\ \emph {et~al.}(2011)\citenamefont {Dunne},
  \citenamefont {Mazza},\ and\ \citenamefont {Coey}}]{dunne_2011}%
  \BibitemOpen
  \bibfield  {author} {\bibinfo {author} {\bibfnamefont {P.}~\bibnamefont
  {Dunne}}, \bibinfo {author} {\bibfnamefont {L.}~\bibnamefont {Mazza}}, \ and\
  \bibinfo {author} {\bibfnamefont {J.~M.~D.}\ \bibnamefont {Coey}},\ }\href
  {\doibase 10.1103/PhysRevLett.107.024501} {\bibfield  {journal} {\bibinfo
  {journal} {Phys. Rev. Lett.}\ }\textbf {\bibinfo {volume} {107}},\ \bibinfo
  {pages} {024501} (\bibinfo {year} {2011})}\BibitemShut {NoStop}%
\bibitem [{\citenamefont {Dunne}\ and\ \citenamefont
  {Coey}(2012)}]{dunne_2012}%
  \BibitemOpen
  \bibfield  {author} {\bibinfo {author} {\bibfnamefont {P.}~\bibnamefont
  {Dunne}}\ and\ \bibinfo {author} {\bibfnamefont {J.~M.~D.}\ \bibnamefont
  {Coey}},\ }\href {\doibase 10.1103/PhysRevB.85.224411} {\bibfield  {journal}
  {\bibinfo  {journal} {Phys. Rev. B}\ }\textbf {\bibinfo {volume} {85}},\
  \bibinfo {pages} {224411} (\bibinfo {year} {2012})}\BibitemShut {NoStop}%
\bibitem [{rem()}]{remark_kelvin}%
  \BibitemOpen
  \href@noop {} {}\bibinfo {note} {Note that this expression is only valid if
  $\mathbf{B}\,{\approx}\, \mu_0 \mathbf{H}$.}\BibitemShut {Stop}%
\bibitem [{\citenamefont {Coey}\ \emph {et~al.}(2009)\citenamefont {Coey},
  \citenamefont {Aogaki}, \citenamefont {Byrne},\ and\ \citenamefont
  {Stamenov}}]{coey_2009}%
  \BibitemOpen
  \bibfield  {author} {\bibinfo {author} {\bibfnamefont {J.~M.~D.}\
  \bibnamefont {Coey}}, \bibinfo {author} {\bibfnamefont {R.}~\bibnamefont
  {Aogaki}}, \bibinfo {author} {\bibfnamefont {F.}~\bibnamefont {Byrne}}, \
  and\ \bibinfo {author} {\bibfnamefont {P.}~\bibnamefont {Stamenov}},\ }\href
  {\doibase 10.1073/pnas.0900561106} {\bibfield  {journal} {\bibinfo  {journal}
  {Proc. Natl. Acad. Sci. U.S.A.}\ }\textbf {\bibinfo {volume} {106}},\
  \bibinfo {pages} {8811} (\bibinfo {year} {2009})}\BibitemShut {NoStop}%
\bibitem [{\citenamefont {Braithwaite}\ \emph {et~al.}(1991)\citenamefont
  {Braithwaite}, \citenamefont {Beaugnon},\ and\ \citenamefont
  {Tournier}}]{braithwaite_1991}%
  \BibitemOpen
  \bibfield  {author} {\bibinfo {author} {\bibfnamefont {D.}~\bibnamefont
  {Braithwaite}}, \bibinfo {author} {\bibfnamefont {E.}~\bibnamefont
  {Beaugnon}}, \ and\ \bibinfo {author} {\bibfnamefont {R.}~\bibnamefont
  {Tournier}},\ }\href@noop {} {\bibfield  {journal} {\bibinfo  {journal}
  {Nature}\ }\textbf {\bibinfo {volume} {354}},\ \bibinfo {pages} {134}
  (\bibinfo {year} {1991})}\BibitemShut {NoStop}%
\bibitem [{\citenamefont {Rodrigues}\ \emph {et~al.}(2019)\citenamefont
  {Rodrigues}, \citenamefont {Lukina}, \citenamefont {Dehaeck}, \citenamefont
  {Colinet}, \citenamefont {Binnemans},\ and\ \citenamefont
  {Fransaer}}]{rodrigues_2019}%
  \BibitemOpen
  \bibfield  {author} {\bibinfo {author} {\bibfnamefont {I.~R.}\ \bibnamefont
  {Rodrigues}}, \bibinfo {author} {\bibfnamefont {L.}~\bibnamefont {Lukina}},
  \bibinfo {author} {\bibfnamefont {S.}~\bibnamefont {Dehaeck}}, \bibinfo
  {author} {\bibfnamefont {P.}~\bibnamefont {Colinet}}, \bibinfo {author}
  {\bibfnamefont {K.}~\bibnamefont {Binnemans}}, \ and\ \bibinfo {author}
  {\bibfnamefont {J.}~\bibnamefont {Fransaer}},\ }\href {\doibase
  10.1021/acs.jpcc.9b06706} {\bibfield  {journal} {\bibinfo  {journal} {J.
  Phys. Chem. C}\ }\textbf {\bibinfo {volume} {123}},\ \bibinfo {pages} {23131}
  (\bibinfo {year} {2019})}\BibitemShut {NoStop}%
\bibitem [{\citenamefont {Yang}\ \emph {et~al.}(2012)\citenamefont {Yang},
  \citenamefont {Tschulik}, \citenamefont {Uhlemann}, \citenamefont
  {Odenbach},\ and\ \citenamefont {Eckert}}]{eckert_2012}%
  \BibitemOpen
  \bibfield  {author} {\bibinfo {author} {\bibfnamefont {X.}~\bibnamefont
  {Yang}}, \bibinfo {author} {\bibfnamefont {K.}~\bibnamefont {Tschulik}},
  \bibinfo {author} {\bibfnamefont {M.}~\bibnamefont {Uhlemann}}, \bibinfo
  {author} {\bibfnamefont {S.}~\bibnamefont {Odenbach}}, \ and\ \bibinfo
  {author} {\bibfnamefont {K.}~\bibnamefont {Eckert}},\ }\href {\doibase
  10.1021/jz301561q} {\bibfield  {journal} {\bibinfo  {journal} {J. Phys. Chem.
  Lett.}\ }\textbf {\bibinfo {volume} {3}},\ \bibinfo {pages} {3559} (\bibinfo
  {year} {2012})}\BibitemShut {NoStop}%
\bibitem [{\citenamefont {Pulko}\ \emph {et~al.}(2014)\citenamefont {Pulko},
  \citenamefont {Yang}, \citenamefont {Lei}, \citenamefont {Odenbach},\ and\
  \citenamefont {Eckert}}]{eckert_2014}%
  \BibitemOpen
  \bibfield  {author} {\bibinfo {author} {\bibfnamefont {B.}~\bibnamefont
  {Pulko}}, \bibinfo {author} {\bibfnamefont {X.}~\bibnamefont {Yang}},
  \bibinfo {author} {\bibfnamefont {Z.}~\bibnamefont {Lei}}, \bibinfo {author}
  {\bibfnamefont {S.}~\bibnamefont {Odenbach}}, \ and\ \bibinfo {author}
  {\bibfnamefont {K.}~\bibnamefont {Eckert}},\ }\href {\doibase
  10.1063/1.4903794} {\bibfield  {journal} {\bibinfo  {journal} {Appl. Phys.
  Lett.}\ }\textbf {\bibinfo {volume} {105}},\ \bibinfo {pages} {232407}
  (\bibinfo {year} {2014})}\BibitemShut {NoStop}%
\bibitem [{\citenamefont {Rodrigues}\ \emph {et~al.}(2017)\citenamefont
  {Rodrigues}, \citenamefont {Lukina}, \citenamefont {Dehaeck}, \citenamefont
  {Colinet}, \citenamefont {Binnemans},\ and\ \citenamefont
  {Fransaer}}]{rodrigues_2017}%
  \BibitemOpen
  \bibfield  {author} {\bibinfo {author} {\bibfnamefont {I.~R.}\ \bibnamefont
  {Rodrigues}}, \bibinfo {author} {\bibfnamefont {L.}~\bibnamefont {Lukina}},
  \bibinfo {author} {\bibfnamefont {S.}~\bibnamefont {Dehaeck}}, \bibinfo
  {author} {\bibfnamefont {P.}~\bibnamefont {Colinet}}, \bibinfo {author}
  {\bibfnamefont {K.}~\bibnamefont {Binnemans}}, \ and\ \bibinfo {author}
  {\bibfnamefont {J.}~\bibnamefont {Fransaer}},\ }\href {\doibase
  10.1021/acs.jpclett.7b02226} {\bibfield  {journal} {\bibinfo  {journal} {J.
  Phys. Chem. Lett.}\ }\textbf {\bibinfo {volume} {8}},\ \bibinfo {pages}
  {5301} (\bibinfo {year} {2017})}\BibitemShut {NoStop}%
\bibitem [{\citenamefont {Lei}\ \emph {et~al.}(2017)\citenamefont {Lei},
  \citenamefont {Fritzsche},\ and\ \citenamefont {Eckert}}]{eckert_2017}%
  \BibitemOpen
  \bibfield  {author} {\bibinfo {author} {\bibfnamefont {Z.}~\bibnamefont
  {Lei}}, \bibinfo {author} {\bibfnamefont {B.}~\bibnamefont {Fritzsche}}, \
  and\ \bibinfo {author} {\bibfnamefont {K.}~\bibnamefont {Eckert}},\ }\href
  {\doibase 10.1021/acs.jpcc.7b07344} {\bibfield  {journal} {\bibinfo
  {journal} {J. Phys. Chem. C}\ }\textbf {\bibinfo {volume} {121}},\ \bibinfo
  {pages} {24576} (\bibinfo {year} {2017})}\BibitemShut {NoStop}%
\bibitem [{\citenamefont {Franczak}\ \emph {et~al.}(2016)\citenamefont
  {Franczak}, \citenamefont {Binnemans},\ and\ \citenamefont
  {Fransaer}}]{franczak_2016}%
  \BibitemOpen
  \bibfield  {author} {\bibinfo {author} {\bibfnamefont {A.}~\bibnamefont
  {Franczak}}, \bibinfo {author} {\bibfnamefont {K.}~\bibnamefont {Binnemans}},
  \ and\ \bibinfo {author} {\bibfnamefont {J.}~\bibnamefont {Fransaer}},\
  }\href {\doibase 10.1039/C6CP02575G} {\bibfield  {journal} {\bibinfo
  {journal} {Phys. Chem. Chem. Phys.}\ }\textbf {\bibinfo {volume} {18}},\
  \bibinfo {pages} {27342} (\bibinfo {year} {2016})}\BibitemShut {NoStop}%
\bibitem [{\citenamefont {Ko{\l}czyk}\ \emph {et~al.}(2016)\citenamefont
  {Ko{\l}czyk}, \citenamefont {Wojnicki}, \citenamefont {Kuty{\l}a},
  \citenamefont {Kowalik}, \citenamefont {{\.Z}abi{\'n}ski},\ and\
  \citenamefont {Cristofolini}}]{kolczyk_2016}%
  \BibitemOpen
  \bibfield  {author} {\bibinfo {author} {\bibfnamefont {K.}~\bibnamefont
  {Ko{\l}czyk}}, \bibinfo {author} {\bibfnamefont {M.}~\bibnamefont
  {Wojnicki}}, \bibinfo {author} {\bibfnamefont {D.}~\bibnamefont {Kuty{\l}a}},
  \bibinfo {author} {\bibfnamefont {R.}~\bibnamefont {Kowalik}}, \bibinfo
  {author} {\bibfnamefont {P.}~\bibnamefont {{\.Z}abi{\'n}ski}}, \ and\
  \bibinfo {author} {\bibfnamefont {A.}~\bibnamefont {Cristofolini}},\
  }\href@noop {} {\bibfield  {journal} {\bibinfo  {journal} {Arch. Metall.
  Mater.}\ }\textbf {\bibinfo {volume} {61}},\ \bibinfo {pages} {1919}
  (\bibinfo {year} {2016})}\BibitemShut {NoStop}%
\bibitem [{\citenamefont {Kolczyk-Siedlecka}\ \emph {et~al.}(2019)\citenamefont
  {Kolczyk-Siedlecka}, \citenamefont {Wojnicki}, \citenamefont {Yang},
  \citenamefont {Mutschke},\ and\ \citenamefont {Zabinski}}]{kolczyk_2019}%
  \BibitemOpen
  \bibfield  {author} {\bibinfo {author} {\bibfnamefont {K.}~\bibnamefont
  {Kolczyk-Siedlecka}}, \bibinfo {author} {\bibfnamefont {M.}~\bibnamefont
  {Wojnicki}}, \bibinfo {author} {\bibfnamefont {X.}~\bibnamefont {Yang}},
  \bibinfo {author} {\bibfnamefont {G.}~\bibnamefont {Mutschke}}, \ and\
  \bibinfo {author} {\bibfnamefont {P.}~\bibnamefont {Zabinski}},\ }\href
  {https://doi.org/10.1007/s41981-019-00039-8} {\bibfield  {journal} {\bibinfo
  {journal} {J. Flow Chem.}\ }\textbf {\bibinfo {volume} {9}},\ \bibinfo
  {pages} {175} (\bibinfo {year} {2019})}\BibitemShut {NoStop}%
\bibitem [{\citenamefont {Noddack}\ and\ \citenamefont
  {Wicht}(1952)}]{noddack_1952}%
  \BibitemOpen
  \bibfield  {author} {\bibinfo {author} {\bibfnamefont {W.}~\bibnamefont
  {Noddack}}\ and\ \bibinfo {author} {\bibfnamefont {E.}~\bibnamefont
  {Wicht}},\ }\href@noop {} {\bibfield  {journal} {\bibinfo  {journal} {Ber.
  Bunsenges. Phys. Chem.}\ }\textbf {\bibinfo {volume} {56}},\ \bibinfo {pages}
  {893} (\bibinfo {year} {1952})}\BibitemShut {NoStop}%
\bibitem [{\citenamefont {Noddack}\ and\ \citenamefont
  {Wicht}(1955)}]{noddack_1955}%
  \BibitemOpen
  \bibfield  {author} {\bibinfo {author} {\bibfnamefont {I.}~\bibnamefont
  {Noddack}}\ and\ \bibinfo {author} {\bibfnamefont {E.}~\bibnamefont
  {Wicht}},\ }\href@noop {} {\bibfield  {journal} {\bibinfo  {journal} {Chem.
  Techn.}\ }\textbf {\bibinfo {volume} {7}},\ \bibinfo {pages} {3} (\bibinfo
  {year} {1955})}\BibitemShut {NoStop}%
\bibitem [{\citenamefont {Noddack}\ \emph {et~al.}(1958)\citenamefont
  {Noddack}, \citenamefont {Noddack},\ and\ \citenamefont
  {Wicht}}]{noddack_1958}%
  \BibitemOpen
  \bibfield  {author} {\bibinfo {author} {\bibfnamefont {W.}~\bibnamefont
  {Noddack}}, \bibinfo {author} {\bibfnamefont {I.}~\bibnamefont {Noddack}}, \
  and\ \bibinfo {author} {\bibfnamefont {E.}~\bibnamefont {Wicht}},\ }\href
  {\doibase 10.1002/bbpc.19580620112} {\bibfield  {journal} {\bibinfo
  {journal} {Ber. Bunsenges. Phys. Chem.}\ }\textbf {\bibinfo {volume} {62}},\
  \bibinfo {pages} {77} (\bibinfo {year} {1958})}\BibitemShut {NoStop}%
\bibitem [{\citenamefont {Kardjilov}\ \emph {et~al.}(2011)\citenamefont
  {Kardjilov}, \citenamefont {Manke}, \citenamefont {Hilger}, \citenamefont
  {Strobl},\ and\ \citenamefont {Banhart}}]{kardjilov_2011}%
  \BibitemOpen
  \bibfield  {author} {\bibinfo {author} {\bibfnamefont {N.}~\bibnamefont
  {Kardjilov}}, \bibinfo {author} {\bibfnamefont {I.}~\bibnamefont {Manke}},
  \bibinfo {author} {\bibfnamefont {A.}~\bibnamefont {Hilger}}, \bibinfo
  {author} {\bibfnamefont {M.}~\bibnamefont {Strobl}}, \ and\ \bibinfo {author}
  {\bibfnamefont {J.}~\bibnamefont {Banhart}},\ }\href {\doibase
  https://doi.org/10.1016/S1369-7021(11)70139-0} {\bibfield  {journal}
  {\bibinfo  {journal} {Mater. Today}\ }\textbf {\bibinfo {volume} {14}},\
  \bibinfo {pages} {248 } (\bibinfo {year} {2011})}\BibitemShut {NoStop}%
\bibitem [{\citenamefont {Perfect}\ \emph {et~al.}(2014)\citenamefont
  {Perfect}, \citenamefont {Cheng}, \citenamefont {Kang}, \citenamefont
  {Bilheux}, \citenamefont {Lamanna}, \citenamefont {Gragg},\ and\
  \citenamefont {Wright}}]{perfect_2014}%
  \BibitemOpen
  \bibfield  {author} {\bibinfo {author} {\bibfnamefont {E.}~\bibnamefont
  {Perfect}}, \bibinfo {author} {\bibfnamefont {C.-L.}\ \bibnamefont {Cheng}},
  \bibinfo {author} {\bibfnamefont {M.}~\bibnamefont {Kang}}, \bibinfo {author}
  {\bibfnamefont {H.~Z.}\ \bibnamefont {Bilheux}}, \bibinfo {author}
  {\bibfnamefont {J.~M.}\ \bibnamefont {Lamanna}}, \bibinfo {author}
  {\bibfnamefont {M.~J.}\ \bibnamefont {Gragg}}, \ and\ \bibinfo {author}
  {\bibfnamefont {D.~M.}\ \bibnamefont {Wright}},\ }\href {\doibase
  https://doi.org/10.1016/j.earscirev.2013.11.012} {\bibfield  {journal}
  {\bibinfo  {journal} {Earth-Sci. Rev}\ }\textbf {\bibinfo {volume} {129}},\
  \bibinfo {pages} {120 } (\bibinfo {year} {2014})}\BibitemShut {NoStop}%
\bibitem [{\citenamefont {Burca}\ \emph {et~al.}(2018)\citenamefont {Burca},
  \citenamefont {Nagella}, \citenamefont {Clark}, \citenamefont {Tasev},
  \citenamefont {Rahman}, \citenamefont {Garwood}, \citenamefont {Spencer},
  \citenamefont {Turner},\ and\ \citenamefont {Kelleher}}]{burca_2018}%
  \BibitemOpen
  \bibfield  {author} {\bibinfo {author} {\bibfnamefont {G.}~\bibnamefont
  {Burca}}, \bibinfo {author} {\bibfnamefont {S.}~\bibnamefont {Nagella}},
  \bibinfo {author} {\bibfnamefont {T.}~\bibnamefont {Clark}}, \bibinfo
  {author} {\bibfnamefont {D.}~\bibnamefont {Tasev}}, \bibinfo {author}
  {\bibfnamefont {I.~A.}\ \bibnamefont {Rahman}}, \bibinfo {author}
  {\bibfnamefont {R.~J.}\ \bibnamefont {Garwood}}, \bibinfo {author}
  {\bibfnamefont {A.~R.~T.}\ \bibnamefont {Spencer}}, \bibinfo {author}
  {\bibfnamefont {M.~J.}\ \bibnamefont {Turner}}, \ and\ \bibinfo {author}
  {\bibfnamefont {J.~F.}\ \bibnamefont {Kelleher}},\ }\href {\doibase
  10.1111/jmi.12761} {\bibfield  {journal} {\bibinfo  {journal} {J. Microsc.}\
  }\textbf {\bibinfo {volume} {272}},\ \bibinfo {pages} {242} (\bibinfo {year}
  {2018})}\BibitemShut {NoStop}%
\bibitem [{\citenamefont {{n{\_}TOF Collaboration}}\ \emph
  {et~al.}(2019)\citenamefont {{n{\_}TOF Collaboration}}, \citenamefont
  {Mastromarco}, \citenamefont {Manna} \emph {et~al.}}]{mastromarco_2019}%
  \BibitemOpen
  \bibfield  {author} {\bibinfo {author} {\bibnamefont {{n{\_}TOF
  Collaboration}}}, \bibinfo {author} {\bibfnamefont {M.}~\bibnamefont
  {Mastromarco}}, \bibinfo {author} {\bibfnamefont {A.}~\bibnamefont {Manna}},
  \emph {et~al.},\ }\href {\doibase 10.1140/epja/i2019-12692-7} {\bibfield
  {journal} {\bibinfo  {journal} {Eur. Phys. J. A}\ }\textbf {\bibinfo {volume}
  {55}},\ \bibinfo {pages} {9} (\bibinfo {year} {2019})}\BibitemShut {NoStop}%
\bibitem [{\citenamefont {Brenizer}(2013)}]{brenizer_2013}%
  \BibitemOpen
  \bibfield  {author} {\bibinfo {author} {\bibfnamefont {J.}~\bibnamefont
  {Brenizer}},\ }\href {\doibase 10.1016/j.phpro.2013.03.002} {\bibfield
  {journal} {\bibinfo  {journal} {Phys. Procedia}\ }\textbf {\bibinfo {volume}
  {43}},\ \bibinfo {pages} {10} (\bibinfo {year} {2013})}\BibitemShut {NoStop}%
\bibitem [{\citenamefont {Trtik}\ \emph {et~al.}(2016)\citenamefont {Trtik},
  \citenamefont {Morgano}, \citenamefont {Bentz},\ and\ \citenamefont
  {Lehmann}}]{trtik_2016}%
  \BibitemOpen
  \bibfield  {author} {\bibinfo {author} {\bibfnamefont {P.}~\bibnamefont
  {Trtik}}, \bibinfo {author} {\bibfnamefont {M.}~\bibnamefont {Morgano}},
  \bibinfo {author} {\bibfnamefont {R.}~\bibnamefont {Bentz}}, \ and\ \bibinfo
  {author} {\bibfnamefont {E.}~\bibnamefont {Lehmann}},\ }\href {\doibase
  10.1016/j.mex.2016.10.001} {\bibfield  {journal} {\bibinfo  {journal}
  {MethodsX}\ }\textbf {\bibinfo {volume} {3}},\ \bibinfo {pages} {535}
  (\bibinfo {year} {2016})}\BibitemShut {NoStop}%
\bibitem [{\citenamefont {Zboray}\ and\ \citenamefont
  {Trtik}(2019)}]{zboray_2019}%
  \BibitemOpen
  \bibfield  {author} {\bibinfo {author} {\bibfnamefont {R.}~\bibnamefont
  {Zboray}}\ and\ \bibinfo {author} {\bibfnamefont {P.}~\bibnamefont {Trtik}},\
  }\href {\doibase https://doi.org/10.1016/j.flowmeasinst.2019.03.005}
  {\bibfield  {journal} {\bibinfo  {journal} {Flow. Meas. Instrum.}\ }\textbf
  {\bibinfo {volume} {66}},\ \bibinfo {pages} {182 } (\bibinfo {year}
  {2019})}\BibitemShut {NoStop}%
\bibitem [{\citenamefont {Ott}\ \emph {et~al.}(2015)\citenamefont {Ott},
  \citenamefont {Loupiac}, \citenamefont {D\'{e}sert}, \citenamefont
  {H\'{e}lary},\ and\ \citenamefont {Lavie}}]{ott_2015}%
  \BibitemOpen
  \bibfield  {author} {\bibinfo {author} {\bibfnamefont {F.}~\bibnamefont
  {Ott}}, \bibinfo {author} {\bibfnamefont {C.}~\bibnamefont {Loupiac}},
  \bibinfo {author} {\bibfnamefont {S.}~\bibnamefont {D\'{e}sert}}, \bibinfo
  {author} {\bibfnamefont {A.}~\bibnamefont {H\'{e}lary}}, \ and\ \bibinfo
  {author} {\bibfnamefont {P.}~\bibnamefont {Lavie}},\ }\href {\doibase
  10.1016/j.phpro.2015.07.009} {\bibfield  {journal} {\bibinfo  {journal}
  {Phys. Procedia}\ }\textbf {\bibinfo {volume} {69}},\ \bibinfo {pages} {67}
  (\bibinfo {year} {2015})}\BibitemShut {NoStop}%
\bibitem [{\citenamefont {Kardjilov}\ \emph {et~al.}(2005)\citenamefont
  {Kardjilov}, \citenamefont {de~Beer}, \citenamefont {Hassanein},
  \citenamefont {Lehmann},\ and\ \citenamefont {Vontobel}}]{kardjilov_2005}%
  \BibitemOpen
  \bibfield  {author} {\bibinfo {author} {\bibfnamefont {N.}~\bibnamefont
  {Kardjilov}}, \bibinfo {author} {\bibfnamefont {F.}~\bibnamefont {de~Beer}},
  \bibinfo {author} {\bibfnamefont {R.}~\bibnamefont {Hassanein}}, \bibinfo
  {author} {\bibfnamefont {E.}~\bibnamefont {Lehmann}}, \ and\ \bibinfo
  {author} {\bibfnamefont {P.}~\bibnamefont {Vontobel}},\ }\href {\doibase
  https://doi.org/10.1016/j.nima.2005.01.159} {\bibfield  {journal} {\bibinfo
  {journal} {Nucl. Instrum. Methods Phys. Res}\ }\textbf {\bibinfo {volume}
  {542}},\ \bibinfo {pages} {336 } (\bibinfo {year} {2005})}\BibitemShut
  {NoStop}%
\bibitem [{\citenamefont {Boillat}\ \emph {et~al.}(2018)\citenamefont
  {Boillat}, \citenamefont {Carminati}, \citenamefont {Schmid}, \citenamefont
  {Gr\"{u}nzweig}, \citenamefont {Hovind}, \citenamefont {Kaestner},
  \citenamefont {Mannes}, \citenamefont {Morgano}, \citenamefont {Siegwart},
  \citenamefont {Trtik}, \citenamefont {Vontobel},\ and\ \citenamefont
  {Lehmann}}]{boillat_2018}%
  \BibitemOpen
  \bibfield  {author} {\bibinfo {author} {\bibfnamefont {P.}~\bibnamefont
  {Boillat}}, \bibinfo {author} {\bibfnamefont {C.}~\bibnamefont {Carminati}},
  \bibinfo {author} {\bibfnamefont {F.}~\bibnamefont {Schmid}}, \bibinfo
  {author} {\bibfnamefont {C.}~\bibnamefont {Gr\"{u}nzweig}}, \bibinfo {author}
  {\bibfnamefont {J.}~\bibnamefont {Hovind}}, \bibinfo {author} {\bibfnamefont
  {A.}~\bibnamefont {Kaestner}}, \bibinfo {author} {\bibfnamefont
  {D.}~\bibnamefont {Mannes}}, \bibinfo {author} {\bibfnamefont
  {M.}~\bibnamefont {Morgano}}, \bibinfo {author} {\bibfnamefont
  {M.}~\bibnamefont {Siegwart}}, \bibinfo {author} {\bibfnamefont
  {P.}~\bibnamefont {Trtik}}, \bibinfo {author} {\bibfnamefont
  {P.}~\bibnamefont {Vontobel}}, \ and\ \bibinfo {author} {\bibfnamefont
  {E.}~\bibnamefont {Lehmann}},\ }\href {\doibase 10.1364/OE.26.015769}
  {\bibfield  {journal} {\bibinfo  {journal} {Opt. Express}\ }\textbf {\bibinfo
  {volume} {26}},\ \bibinfo {pages} {15769} (\bibinfo {year}
  {2018})}\BibitemShut {NoStop}%
\bibitem [{\citenamefont {Carminati}\ \emph {et~al.}(2019)\citenamefont
  {Carminati}, \citenamefont {Boillat}, \citenamefont {Schmid}, \citenamefont
  {Vontobel}, \citenamefont {Hovind}, \citenamefont {Morgano}, \citenamefont
  {Raventos}, \citenamefont {Siegwart}, \citenamefont {Mannes}, \citenamefont
  {Gr\"{u}nzweig}, \citenamefont {Trtik}, \citenamefont {Lehmann},
  \citenamefont {Strobl},\ and\ \citenamefont {Kaestner}}]{carminati_2019}%
  \BibitemOpen
  \bibfield  {author} {\bibinfo {author} {\bibfnamefont {C.}~\bibnamefont
  {Carminati}}, \bibinfo {author} {\bibfnamefont {P.}~\bibnamefont {Boillat}},
  \bibinfo {author} {\bibfnamefont {F.}~\bibnamefont {Schmid}}, \bibinfo
  {author} {\bibfnamefont {P.}~\bibnamefont {Vontobel}}, \bibinfo {author}
  {\bibfnamefont {J.}~\bibnamefont {Hovind}}, \bibinfo {author} {\bibfnamefont
  {M.}~\bibnamefont {Morgano}}, \bibinfo {author} {\bibfnamefont
  {M.}~\bibnamefont {Raventos}}, \bibinfo {author} {\bibfnamefont
  {M.}~\bibnamefont {Siegwart}}, \bibinfo {author} {\bibfnamefont
  {D.}~\bibnamefont {Mannes}}, \bibinfo {author} {\bibfnamefont
  {C.}~\bibnamefont {Gr\"{u}nzweig}}, \bibinfo {author} {\bibfnamefont
  {P.}~\bibnamefont {Trtik}}, \bibinfo {author} {\bibfnamefont
  {E.}~\bibnamefont {Lehmann}}, \bibinfo {author} {\bibfnamefont
  {M.}~\bibnamefont {Strobl}}, \ and\ \bibinfo {author} {\bibfnamefont
  {A.}~\bibnamefont {Kaestner}},\ }\href {\doibase
  10.1371/journal.pone.0210300} {\bibfield  {journal} {\bibinfo  {journal}
  {PLOS ONE}\ }\textbf {\bibinfo {volume} {14}},\ \bibinfo {pages} {1}
  (\bibinfo {year} {2019})}\BibitemShut {NoStop}%
\bibitem [{\citenamefont {Furlani}(2001)}]{furlani_2001}%
  \BibitemOpen
  \bibfield  {author} {\bibinfo {author} {\bibfnamefont {E.~P.}\ \bibnamefont
  {Furlani}},\ }\href@noop {} {\emph {\bibinfo {title} {Permanent magnet and
  electromechanical devices: materials, analysis, and applications}}}\
  (\bibinfo  {publisher} {Academic Press},\ \bibinfo {address} {San Diego},\
  \bibinfo {year} {2001})\ pp.\ \bibinfo {pages} {208--217}\BibitemShut
  {NoStop}%
\bibitem [{\citenamefont {Cussler}(2009)}]{cussler_2009}%
  \BibitemOpen
  \bibfield  {author} {\bibinfo {author} {\bibfnamefont {E.~L.}\ \bibnamefont
  {Cussler}},\ }\href@noop {} {\emph {\bibinfo {title} {Diffusion: Mass
  Transfer in Fluid Systems}}}\ (\bibinfo  {publisher} {Cambridge University
  Press},\ \bibinfo {year} {2009})\ p.\ \bibinfo {pages} {162}\BibitemShut
  {NoStop}%
\bibitem [{\citenamefont {Sears}(1992)}]{varley_1992}%
  \BibitemOpen
  \bibfield  {author} {\bibinfo {author} {\bibfnamefont {V.~F.}\ \bibnamefont
  {Sears}},\ }\href {\doibase 10.1080/10448639208218770} {\bibfield  {journal}
  {\bibinfo  {journal} {Neutron News}\ }\textbf {\bibinfo {volume} {3}},\
  \bibinfo {pages} {26} (\bibinfo {year} {1992})}\BibitemShut {NoStop}%
\bibitem [{\citenamefont {Huang}\ \emph {et~al.}(2007)\citenamefont {Huang},
  \citenamefont {De~Luca}, \citenamefont {Atherton}, \citenamefont {Bird},
  \citenamefont {Rosenblatt},\ and\ \citenamefont {Carl\`es}}]{huang_2007}%
  \BibitemOpen
  \bibfield  {author} {\bibinfo {author} {\bibfnamefont {Z.}~\bibnamefont
  {Huang}}, \bibinfo {author} {\bibfnamefont {A.}~\bibnamefont {De~Luca}},
  \bibinfo {author} {\bibfnamefont {T.~J.}\ \bibnamefont {Atherton}}, \bibinfo
  {author} {\bibfnamefont {M.}~\bibnamefont {Bird}}, \bibinfo {author}
  {\bibfnamefont {C.}~\bibnamefont {Rosenblatt}}, \ and\ \bibinfo {author}
  {\bibfnamefont {P.}~\bibnamefont {Carl\`es}},\ }\href {\doibase
  10.1103/PhysRevLett.99.204502} {\bibfield  {journal} {\bibinfo  {journal}
  {Phys. Rev. Lett.}\ }\textbf {\bibinfo {volume} {99}},\ \bibinfo {pages}
  {204502} (\bibinfo {year} {2007})}\BibitemShut {NoStop}%
\bibitem [{\citenamefont {Tsiklashvili}\ \emph {et~al.}(2012)\citenamefont
  {Tsiklashvili}, \citenamefont {Colio}, \citenamefont {Likhachev},\ and\
  \citenamefont {Jacobs}}]{tsiklashvili_2012}%
  \BibitemOpen
  \bibfield  {author} {\bibinfo {author} {\bibfnamefont {V.}~\bibnamefont
  {Tsiklashvili}}, \bibinfo {author} {\bibfnamefont {P.~E.~R.}\ \bibnamefont
  {Colio}}, \bibinfo {author} {\bibfnamefont {O.~A.}\ \bibnamefont
  {Likhachev}}, \ and\ \bibinfo {author} {\bibfnamefont {J.~W.}\ \bibnamefont
  {Jacobs}},\ }\href {\doibase 10.1063/1.4721898} {\bibfield  {journal}
  {\bibinfo  {journal} {Phys. Fluids}\ }\textbf {\bibinfo {volume} {24}},\
  \bibinfo {pages} {052106} (\bibinfo {year} {2012})}\BibitemShut {NoStop}%
\bibitem [{\citenamefont {Turner}\ and\ \citenamefont
  {Stommel}(1964)}]{turner_1964}%
  \BibitemOpen
  \bibfield  {author} {\bibinfo {author} {\bibfnamefont {J.~S.}\ \bibnamefont
  {Turner}}\ and\ \bibinfo {author} {\bibfnamefont {H.}~\bibnamefont
  {Stommel}},\ }\href@noop {} {\bibfield  {journal} {\bibinfo  {journal} {Proc.
  Natl. Acad. Sci. U.S.A.}\ }\textbf {\bibinfo {volume} {52}},\ \bibinfo
  {pages} {49} (\bibinfo {year} {1964})}\BibitemShut {NoStop}%
\bibitem [{\citenamefont {Stern}\ and\ \citenamefont
  {Turner}(1969)}]{stern_1969}%
  \BibitemOpen
  \bibfield  {author} {\bibinfo {author} {\bibfnamefont {M.~E.}\ \bibnamefont
  {Stern}}\ and\ \bibinfo {author} {\bibfnamefont {J.~S.}\ \bibnamefont
  {Turner}},\ }\href {\doibase https://doi.org/10.1016/0011-7471(69)90038-2}
  {\bibfield  {journal} {\bibinfo  {journal} {Deep-Sea Res. Oceanogr. Abstr.}\
  }\textbf {\bibinfo {volume} {16}},\ \bibinfo {pages} {497 } (\bibinfo {year}
  {1969})}\BibitemShut {NoStop}%
\bibitem [{\citenamefont {Taylor}\ and\ \citenamefont
  {Veronis}(1986)}]{taylor_1986}%
  \BibitemOpen
  \bibfield  {author} {\bibinfo {author} {\bibfnamefont {J.}~\bibnamefont
  {Taylor}}\ and\ \bibinfo {author} {\bibfnamefont {G.}~\bibnamefont
  {Veronis}},\ }\href {\doibase 10.1126/science.231.4733.39} {\bibfield
  {journal} {\bibinfo  {journal} {Science}\ }\textbf {\bibinfo {volume}
  {231}},\ \bibinfo {pages} {39} (\bibinfo {year} {1986})}\BibitemShut
  {NoStop}%
\bibitem [{\citenamefont {Morgano}\ \emph {et~al.}(2015)\citenamefont
  {Morgano}, \citenamefont {Lehmann},\ and\ \citenamefont
  {Strobl}}]{morgano_2015}%
  \BibitemOpen
  \bibfield  {author} {\bibinfo {author} {\bibfnamefont {M.}~\bibnamefont
  {Morgano}}, \bibinfo {author} {\bibfnamefont {E.}~\bibnamefont {Lehmann}}, \
  and\ \bibinfo {author} {\bibfnamefont {M.}~\bibnamefont {Strobl}},\ }\href
  {\doibase https://doi.org/10.1016/j.phpro.2015.07.022} {\bibfield  {journal}
  {\bibinfo  {journal} {Phys. Procedia}\ }\textbf {\bibinfo {volume} {69}},\
  \bibinfo {pages} {152 } (\bibinfo {year} {2015})}\BibitemShut {NoStop}%
\bibitem [{\citenamefont {Trtik}\ and\ \citenamefont
  {Lehmann}(2016)}]{trtik_2016_1}%
  \BibitemOpen
  \bibfield  {author} {\bibinfo {author} {\bibfnamefont {P.}~\bibnamefont
  {Trtik}}\ and\ \bibinfo {author} {\bibfnamefont {E.~H.}\ \bibnamefont
  {Lehmann}},\ }\href {\doibase 10.1088/1742-6596/746/1/012004} {\bibfield
  {journal} {\bibinfo  {journal} {J. Phys. Conf. Ser.}\ }\textbf {\bibinfo
  {volume} {746}},\ \bibinfo {pages} {012004} (\bibinfo {year}
  {2016})}\BibitemShut {NoStop}%
\end{thebibliography}%

\onecolumngrid
\newpage

\begin{center}
  \textbf{\large Supplementary Material: \\[.2cm]Neutron imaging of liquid-liquid systems containing paramagnetic salt solutions}\\[.3cm]
  T. A. Butcher,$^{1,*}$ G. J. M. Formon,$^{2}$ P. Dunne$^2$, T. M. Hermans$^2$, F. Ott$^3$, L. Noirez$^3$, and J. M. D. Coey$^1$\\[.2cm]
  {\itshape ${}^1$School of Physics and CRANN, Trinity College, Dublin 2, Ireland\\
  ${}^2$Universit\'{e} de Strasbourg, CNRS, ISIS UMR 7006, 67000 Strasbourg, France\\
  ${}^3$Laboratoire L\'eon Brillouin (CEA-CNRS), Universit\'{e} Paris-Saclay, CEA-Saclay, 91191 Gif-sur-Yvette, France\\}
  ${}^*$tbutcher@tcd.ie\\
(Dated: December 19, 2019)\\[1cm]
\end{center}

\setcounter{equation}{0}
\setcounter{figure}{0}
\setcounter{table}{0}
\setcounter{page}{1}
\setcounter{section}{0}
\renewcommand{\theequation}{S\arabic{equation}}
\renewcommand{\thefigure}{S\arabic{figure}}
\renewcommand{\bibnumfmt}[1]{[S#1]}
\renewcommand{\citenumfont}[1]{S#1}
\renewcommand{\thesection}{S\arabic{section}}
\renewcommand{\thesubsection}{S\arabic{subsection}}
\thispagestyle{empty}

\section*{Additional Time sequenced neutron images}
\subsection{Unhindered mixing by salt fingering}
\begin{figure}[h]
	\centering
	\includegraphics[width=0.4\columnwidth]{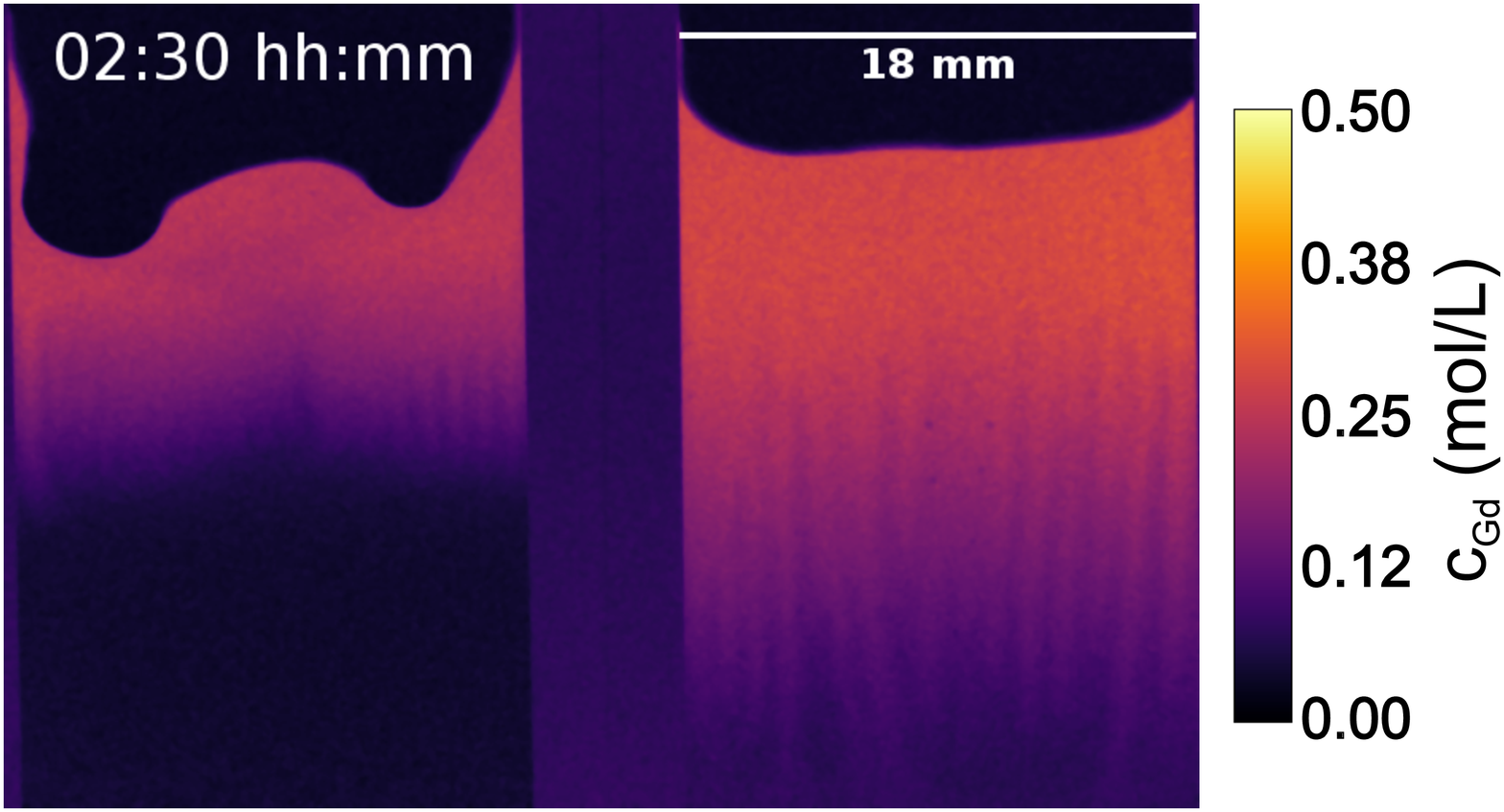}
	\caption{Two 1\,mm path length quartz cuvettes with 75\,\textmu L 0.4\,M \Gd (left) and 100\,\textmu L 0.5\,M \Gd (right) above 300\,\textmu L of 1.3\,M \Y. \Gd solutions were injected above the \Y solution surface in the left cuvette and below it in the right cuvette. Salt fingers form in both systems and mix the solutions within 7\,h. Left cuvette: The surface profile is caused by capillary forces. The salt fingers begin to form at the side of the cuvette and propagate inwards. (Multimedia view)}
	\label{supl:salt_fingers}
\end{figure}	

\subsection{Magnetic confinement of a drop of \Gd solution}
\begin{figure}[h]
	\centering
	\includegraphics[width=0.85\columnwidth]{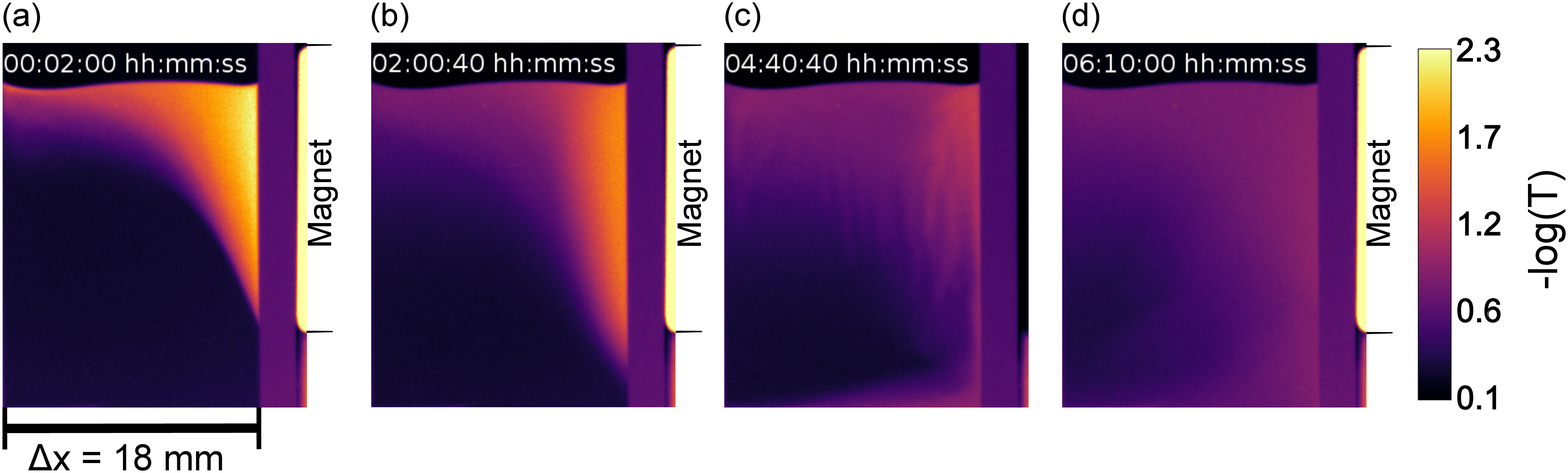}
	\caption{50\,\textmu L 0.5\,M \Gd solution (\mbox{$\rho = 1200\,\mathrm{kg}\,\mathrm{m^{-3}}$}) above 800\,\textmu L 1.3\,M \Y solution (\mbox{$\rho = 1320\,\mathrm{kg}\,\mathrm{m^{-3}}$}) in a 2\,mm path length quartz cuvette. The acquisition time was 80\,s for this measurement. The calibration from Fig~2 is not valid for path lengths above 1\,mm. Therefore, the negative of the logarithm of the transmittance is shown. (a) The \Gd drop migrates to the cubic 20\,mm magnet at the side of the cuvette within 2\,min. (b) The trapped drop in the magnetic field gradient (\mbox{see Fig.~1(b)}) gradually diffuses into the \Y solution over the course of 4.5\,h. (c) Salt fingers appear immediately after removal of the magnet at $\Delta t=4$\,h\,$30$\,min. (d) At $\Delta t=6$\,h\,$10$\,min, the stratification disappears after placing the magnet next to the cuvette again. (Multimedia view)}
	\label{supl:trap}
\end{figure}	

%

\end{document}